\documentclass[%
 reprint,
 amsmath,amssymb,
 aps,
]{revtex4-2}
\usepackage{hyperref}
\usepackage{multirow}
\usepackage{float}
\usepackage{colortbl}
\usepackage{xcolor}
\usepackage{graphicx}
\usepackage{dcolumn}
\usepackage{bm}

\usepackage{tabularx}
\begin{document}

\preprint{APS/123-QED}

\title{Noise-induce coexisting firing patterns in  hybrid-synaptic interacting networks }

\author{Xinyi Wang$^1$}
\author{Xiyun  Zhang$^2$}
\author{Muhua Zheng$^1$}
\author{Leijun  Xu$^3$}
\author{Kesheng Xu$^1$}
\email{ksxu@ujs.edu.cn}
\affiliation{%
1.School of physics and Electronic Engineering, Jiangsu University,Zhenjiang,Jiangsu,212013,China \\
2. Department of Physics, Jinan University, Guangzhou Guangdong, 510632,China\\
3.School of electrical and  information engineering,Jiangsu University, Zhenjiang,Jiangsu,212013, China }

\date{\today}

\begin{abstract}
Synaptic noise plays a major role in setting up coexistence of various firing patterns, but the precise mechanisms whereby these synaptic noise  contributes to coexisting firing activities are subtle and remain elusive. To investigate these mechanisms,  neurons with  hybrid synaptic interaction in a  balanced neuronal networks have been recently put forward.  Here we show that both synaptic  noise intensity and excitatory weights can make a greater contribution than variance of synaptic noise to the coexistence of firing states  with slight modification parameters. The resulting  statistical analysis of both  voltage trajectories and their spike trains  reveals two forms of coexisting  firing patterns: time-varying   and parameter-varying multistability. The emergence of  time-varying multistability  as a format  of metstable  state   has been observed  under suitable  parameters settings  of  noise intensity and excitatory  synaptic weight.   While the parameter-varying multistability  is accompanied by coexistence of synchrony state and metastable  (or asynchronous firing state) with slightly varying noise intensity and excitatory weights. Our results offer a series of precise statistical  explanation of the intricate effect of synaptic noise  in neural multistability. This reconciles previous theoretical and numerical works, and confirms the suitability of various statistical methods to investigate multistability in  a hybrid synaptic interacting neuronal networks.

\end{abstract}

\maketitle

\section{INTRODUCTION}
Multistability or coexistence of several possible final stable states (different attractors) in the nervous system has attracted a increased interest, stemming both from new experimental methods for identifying it and from a growing body of modeling work demonstrating its functional consequence\cite{pisarchik2014control}. Many of these studies over several decades have outlined the sources and impact of biological multistability to a better understanding emergence or any dysfunction of brain behavior\cite{cole2016activity}, in particular, associative memory storage, pattern recognition  and the pathophysiology of human diseases\cite{chopek2019multistable} in both living and artificial neural systems. Multistability have long been of interest, but there still much to be known about their underlying mechanisms in neural systems.

Multistability can arise from two possible sources\cite{orio2018chaos,piccinini2021noise}. The first source is from the deterministic chaos of the neural system. For example, it is generally accepted that inherent instability nature of chaos in neural systems, facilitates the  extremely sensitivity to initial condition, to create a rich  varity of coexistence of firing patterns. Deterministic chaos also can induce a definite switching between different coexisting states, both at the level of individual neurons and neural networks.The alternative source of multistability is noise. Some studies demonstrate that noise not only induce multistability, but also drive  transitions between two or more multistable attractors, in systems of coupled oscillators \cite{kim1997multistability} and in a field-dependent relaxation model\cite{buceta2004comprehensive}.Whereas previous studies have focused on neuronal multistability caused by deterministic  chaos or stochastic noise in networks of neurons solely connected by chemical or electrical synapses, we focus here on work mainly relating to noise-driven coexisting firing patterns in a hybrid synaptic interacting balanced neural networks.

    Biologically relevant sources of noise permeating every level of the nervous system from the perception of sensory signals to the generation of motor responses, have beed evidenced by experimental data (see review\cite{faisal2008noise,mcdonnell2011benefits}). For external sensory stimuli to brain, all forms of perception such as chemical sensing and vision are affected by thermodynamic noise, which is namely  sensory noise. Sources of sensory noise include  intrinsically noisy from external sensory stimuli and transducer noise that is generated during the amplification process\cite{mcdonnell2011benefits,lillywhite1979transducer}. It is well known that neuronal activity is intrinsically irregular in the generation of action potentials, their axonal propagation, network interactions, and the synaptic transmission that follows.  This is due to the stochastic opening and closing of ion channels,the constant bombardment of synaptic inputs and synaptic unreliability\cite{rusakov2020noisy}. This type of neuronal noise is cellular noise, which can be of channel noise, synaptic noise and network interactions\cite{faisal2008noise,mcdonnell2011benefits}. Multiple factors contribute to cellular noise, including changes in the interal states  of neurons and networks or random processes inside neurons and neuronal networks.  At end, experimental evidence also suggests the nervous system has to act in the presence of noise in sensing, information processing and
movement in the behavioural task, such as catching a ball. This type sources of neural variability in the force generated by motor neurons and muscle fibres is motor noise\cite{hamilton2004scaling}. While observation of neural noise has been explored in multiple experimental studies, it is not yet well understood the diverse roles of noise in neural computation and brain function based on neuronal theories and models. 

Noise commonly assumed to be a nuisance, and nervous systems develop strategies to filter it. Besides, noise can also induce new organized behaviors in systems that lack in deterministic conditions(see  reviews \cite{ faisal2008noise,mcdonnell2011benefits,rusakov2020noisy}).  Several strategies have been adopted to use noise in this fashion. Most notably,in spike-generating type neurons,  noise can transform threshold nonlinearities by making subthreshold inputs more likely to cross the threshold,  and thus this is  more easy to  generate a spike. This facilitates spike initiation and can improve neural-network behaviour\cite{anderson2000contribution}.  In addtion, studies of both experiments and theoretical models revealed that a subthreshold sensory signal has a better chance of being detected when noise is added, broadly known as stochastic resonance\cite{stocks2000suprathreshold,mcdonnell2011benefits}.  For a small amount of noise, the sensory signal does not cause the system to cross the threshold and few signals are detected. At large noise levels, the response is dominated by the noise. For intermediate noise intensities, however, the noise allows the signal to reach the threshold for detection but does not swamp it. Therefore, stochastic resonance has been evidenced to enhance processing both in theoretical models
of neural systems and in experimental neuroscience. Moreover, neuronal networks that have formed in the presence of noise will be more robust and explore more states, which will facilitate learning and adaptation to the changing demands of a
dynamic environment. In all, the present of noise leading to spontaneous order in neural systems include stochastic resonance, noise-induced phase transitions, and noise-induced bistability. Nevertheless, the contribution of noise to coexisting firing patterns and stochastic switching between these
attractor basins in the banance neural systems have not been systematically explored yet.

Computational models studies show that itinerant dynamics can be basically  to uncover  mechanisms  of coexisting attractor basins and stochastic switching between multistable states. Several dynamical approaches support itinerant dynamics including chaotic itinerancy, heteroclinic cycling, and multistable switching\cite{pisarchik2014control,orio2018chaos,miller2016itinerancy}. Chaotic itinerancy and heteroclinic cycling focus on deterministic dynamics, in which either a chaotic attractor or a series of saddle points connected by heteroclinic orbits allow the system not to settle in a attractor (or saddle) but instead visit one after the other. Thus, chaotic deterministic trajectories is first possible sources of itinerant dynamics. It is well
known that chaos can emerge in insolate neurons or complex neural systems in different scales of networks size\cite{tang2011synchronization,xu2017hyperpolarization,xu2018synchronization}.  On the other hand, multistable switching implies the coexistence of multiple stable attractors.Thus, neural noise is second possible sources of itinerant dynamics,causing switching between different attractors.

All of these previous studies only identify the essential mechanisms for noise,chaos  or  comparation of them driven multistability of networks sole connected by electrical or chemical synapses. In this work, our main motivation is to identify the mechanisms of noise-induced coexisting firing patterns in a hybrid synaptic interacting balanced neuronal networks -- combination  connection and effection of electrical and chemical synapses.  The excitatory population of balanced network is communicated through chemical synapses in a way of small-world topology, and adjacent excitatory cells of small-world neural network is also connected  by electrical synapses. The nodes in our network are the Wang-Buzsaki model as a general model of mammalian neuronal excitability\cite{wang1996gamma}. In case of fixed synaptic noise intensity, our previous research showed that the existence of electrical synaptic connections to excitatory population can cause various firing patterns of interest by slightly changing the chemical synaptic weights\cite{xu2021diversity}.However,the effection of noise together with other facts with complexity, such as synaptic weights,types of synapses and so on, to neural firing pattern in this balanced network have not been well investigated.

In this paper, we study the emergence of coexisting  firing patterns in an excitatory-inhibitory (E/I) balanced network, combining the effect of synaptic noise and chemical connections. We report that there are two typical forms of coexisting firing patterns obeserved in this balanced networks, that are time-varying coexistence and parameter-varying coexistence. The time-varying coexisting firing patterns  is typically metstable state, normally coexistence of synchronous state and traveling wave or alternation between those two states. The parameter-varying coexistence  in this balanced neural networks  is  accompanied by coexistence of coherence and incoherence  firing state with slightly varying control parameters.  With further investigation, we find that  the emergence of  time-varying coexistence  has been observed  under suitable  parameters settings  of  noise intensity and excitatory  synaptic weight, while latter case, the coexisting  neuronal behaviour  is a result of  systems parameters varying,  since neural network  can  display  a single collective behaviour or metstable firing  state for fixed values of systems parameters. Our findings imply that   synaptic  noise intensity and excitatory weights can make a greater contribution than other factors, such as variance of synaptic noise, to the coexistence of various  firing states  with slight modification parameters.

\section{Models and methods}

\textbf{Wang-Buzsáki model.} The Wang-Buzsaki (WB) model\cite{wang1996gamma,calim2018chimera}resembles the dynamics of fast-spiking neurons in the cortex and hippocampus, and it is used here only as a general model of mammalian neuronal excitability. The Wang-Buzsáki model has three states variables for each nodes: membrane potential $V$, the  variable $h$, $n$ corresponding to the spikes-generating $Na^+ $ and $K^+$ voltage-dependent ion currents. The variable $m$ is considered to be instantaneous. The neuronal dynamics of each node can be described as :
\begin{align}\label{WB:isolateneurons}
	C_m\frac{dV}{dt} &=-I_{Na}-I_{K}-I_{L}-I_{Syn}+I_{app} \nonumber \\
	\frac{dh}{dt} &= \phi(\alpha_h(1-h)-\beta_hh) \nonumber \\
	\frac{dn}{dt} &= \phi(\alpha_n(1-n)-\beta_nn) 
	\end{align}
	
 Where $I_L  = g_L(V-E_L)$, $I_{Na} = g_{Na}m^{3}_{\infty}h(V-E_{Na})$  and  $I_K = g_K n^{4}(V-E_K)$ represent the leak currents, transient sodium currents, and the delayed rectifier currents. $I_{syn}$ stands for the synaptic currents and $I_{app}$ (=0 in our simulations) is the injected currents (in $\mu A/cm^2$). The parameters $g_L$, $g_{Na}$, $g_K$ are the maximal conductance density, $E_L$, $E_{Na}$, $E_K$ are the reversal potential and function $m_{\infty}$ is the steady-state activation variables $m$ of the Hodgkin-Huxley type\cite{hodgkin1952quantitative}. The  default value of these parameters and functions are shown in Table \ref{S1_Table}.

\begin{figure}[htp]	
\includegraphics[width=\linewidth]{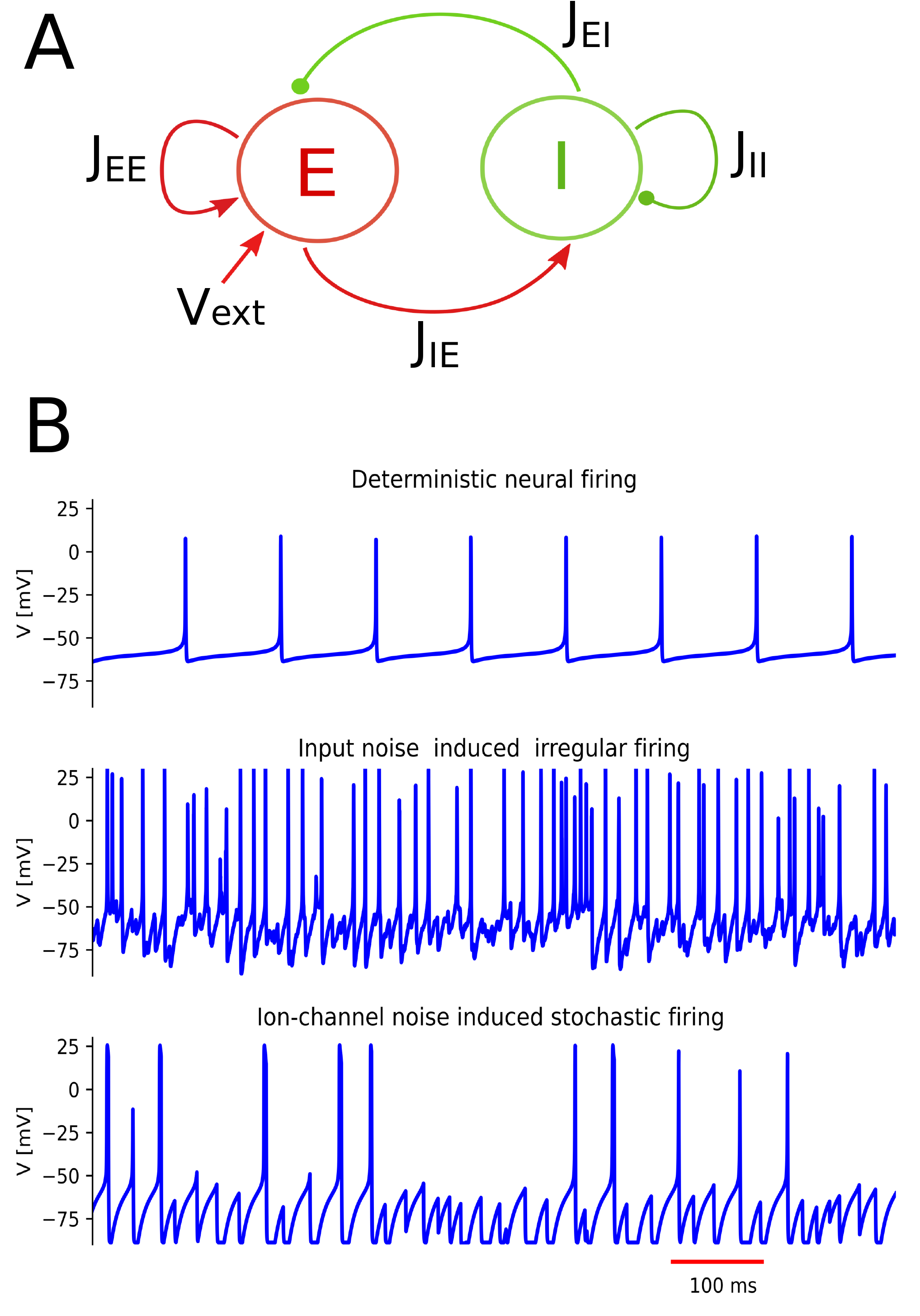}	 
	\caption{ Wang-Buzsáki balance neural network and voltage time course of isolated deterministic and stochastic WB neuron. \textbf{A}. The balanced neural networks containing excitatory (E) and inhibitory ({I}) population . The synaptic communications between these two populations in the way of all-to-all interconnects.  $J_{IE}$ and $J_{EI}$  denote corresponding weights of excitatory and inhibitory synaptic inputs crossing different populations.  $J_{EE}$ and $J_{II}$ present the weights of self-excitatory (in a small world topology) and self-inhibitory connections (in a way of all-to-all topology).     $\nu_{ext}$ is the external excitatory drive to each neuron in the excitatory population.  $J_{gap}$ is weight of electrical synaptic coupling among adjacent neurons in the E-population (not shown here detailly, see Fig.1 in Ref\cite{xu2021diversity}). We fix $J_{II} = 0.04$, $J_{IE} = 0.01$. \textbf{B}. Top: Action potential of isolated deterministic WB neuron when $I_{app}=0.2$; Middle:  irregular neural firing of  a WB neuron induced by Gaussian white-noise when $I_{app}=\mu+\sigma\eta(t)$, where $\mu$ and $\sigma$ are, respectively, the mean and SD of the process and $\eta(t)$ is a Gaussian white-noise variable; Bottom: stochastic neural firing caused by ion channel noise, given by $g_{Kstoch} = g_K*No/Nk$ (note that $Nk=3$ and  $No$  are the number of K+ channels and open channels at equilibrium state).}
	\label{fig:Schematic_figure}
\end{figure}

 To better understand and exhibit effects of various types of noise on WB neuron, three examples of voltage time courses of both deterministic and stochastic WB neurons have well been replotted in \hyperref[fig:Schematic_figure]{Fig.\ref{fig:Schematic_figure}B}. Deterministic WB model, \hyperref[WB:isolateneurons]{Eq.(\ref{WB:isolateneurons})}, has been successfully applied to describe the dynamics of periodic regular firing for mammalian neuronal excitability in \hyperref[fig:Schematic_figure]{Fig.\ref{fig:Schematic_figure}B,Top}. With further  consideration of input noise to WB neuron, such as sensory noise or external input noise,  isolated WB neuron can generate higher frequecy but irregular spikes (in \hyperref[fig:Schematic_figure]{Fig.\ref{fig:Schematic_figure}B,mid}) when $I_{app}=\mu+\sigma\eta(t)$, where $\mu$ and $\sigma$ are, respectively, the mean and SD of the process and $\eta(t)$ is a Gaussian white-noise variable. More importantly, many previous studies showed that neuronal activity is intrinsically irregular in the generation of action potentials due to channel noise, synaptic noise and network interactions\cite{rusakov2020noisy,faisal2008noise,mcdonnell2011benefits}. For ease of comparison with the previous irregular firing, example of intrinsically irregular firing induced by channel noise $g_{Kstoch} = g_K*No/Nk$ (note that $Nk=3$ and  $No$  are the number of K+ channels and open channels at equilibrium state), has been shown clearly in \hyperref[fig:Schematic_figure]{Fig.\ref{fig:Schematic_figure}B,bottom.} As shown in \hyperref[fig:Schematic_figure]{Fig.\ref{fig:Schematic_figure}B,} the former irregular firing impacted  by input noise of continuous-time stochastic processes as Langevin equation, however, the later intrinsically stochastic firing affeced by channel noise in a discrete time set-up of specific open channels. The theoretical underlying mechanisms of generating these significant stochastic neural firing, such as dynamical bifucation, have been so deeply investgated\cite{izhikevich2007dynamical,laing2009stochastic}. This study is to find and ascertain the effects of  intrinsically stochastic of interest on collective behaviour (e.g coexisting firing patterns) of balance neural networks.
 
\textbf{Synaptic dynamics.} The neurons in the networks are connected  both by electrical and chemical synapses. $I_{SynE,i}$ ,    $ I_{SynI,i}$ stand for  total synaptic input currents into neuron $i$ for excitatory population and inhibitory population, respectively, given by:
\begin{widetext}
\begin{align}\label{eq:excSynaCurrents}
I_{SynE,i} = J_{gap} \vec{C}_{gap}^{i} \cdot \vec{D}^{i}+ J_{EE}\vec{L}_{EE}^{i}\cdot \vec{g}_{syn}(V_{i}-E_{synE}) 
+ J_{EI}\vec{L}_{EI}^{i}\cdot \vec{g}_{syn}(V_{i}-E_{synI})+ \sum_{j=1}^{N_{E}}g_{syn,j}(V_{i}-E_{synE}) 
\end{align}

\begin{align}\label{eq:inhiSyna}
I_{SynI,i}  =  J_{II}\vec{L}_{II}^{i}\cdot \vec{g}_{syn}(V_{i}-E_{synI}) + J_{IE} \vec{L}_{IE}^{i}\cdot \vec{g}_{syn}(V_{i}-E_{synE})
\end{align}
\end{widetext}

 The dot products in \hyperref[eq:excSynaCurrents]{Eq.(\ref{eq:excSynaCurrents})} are electrical synaptic inputs, self-excitatory synaptic inputs, inhibitory synaptic inputs, and last term of \hyperref[eq:excSynaCurrents]{Eq.(\ref{eq:excSynaCurrents})} is external excitatory synaptic inputs. The dot products in \hyperref[eq:inhiSyna]{Eq.(\ref{eq:inhiSyna})} represent self-inhibitory synaptic inputs and excitatory synaptic inputs.The vector $\vec{C}_{gap}^{i}=({c}_{gap}^{i1},{c}_{gap}^{i2},\cdots,{c}_{gap}^{ij})$  and $\vec{D}^{i}=(V_{i}-V_{1},V_{i}-V_{2},\cdots,V_{i}-V_{j})$ stand for connection vectors and their diffusive coupling induced by electrical synapses. The elements in vector $\vec{g}_{syn}=(g_{syn,1},g_{syn,2},\cdots,g_{syn,j})$ are chemical synaptic conductances.    The elements of vector $\vec{L}_{mn}^{i} = (l_{mn}^{i1},l_{mn}^{i2},\cdots,l_{mn}^{ij})$ denote the connections between the $i$th neurons of the $m= E,I$ (excitatory, $E$ or inhibtory  $I$) population and the $jth$ neurons of the $n= E,I$ population. More detailed explanations are that: $l_{EE}^{ij}$ and  $l_{II}^{ij}$ are elements of self-excitatory  and self-inhibitory  connections, whereas $l_{IE}^{ij}$ (or $l_{EI}^{ij}$) is element of the  all-to-all matrices of excitatory-to-inhibitory connections (or vice versa). $C_{gap}^{ij}$ is element of electrical synaptic connections in excitatory population. The parameters $J_{EE},J_{II},J_{IE},J_{EI},J_{gap}$ indicate the corresponding synaptic weights of connections $l_{EE}^{ij},l_{II}^{ij},l_{IE}^{ij},l_{EI}^{ij},C_{gap}^{ij}$. Here $L_{mn}^{ij} = 1$ (or $ 0$) is (or not) connections (also see Ref\cite{xu2021diversity}). The  default network parameters used here  are shown in Table \ref{S2_Table}.The updating rule of chemical synaptic conductances,$g_{syn}$, has been described in \hyperref[syn:updating]{Appendix A.\ref{syn:updating}}.

\textbf{Network connectivity.} We consider a balanced neural network, consisting of $N_E= 1000$ excitatory  and  $N_I =250 $ inhibitory neurons  for this paper shown in \hyperref[fig:Schematic_figure]{Fig.\ref{fig:Schematic_figure}}. The excitatory population itself is connected by excitatory synapses in small world topology and its adjacent neurons are inter-connected by gap junctions. The inhibitory population itself is only with all-to-all interaction by
inhibitory synapses. The small-world topology for excitatory population is implemented as two basic steps of the standard algorithm (See Ref.\cite{Watts1998}).

\textbf{Network Activity Characterization.} In this paper we first use the synchronization index, $\chi$, to account for the synchronization level of the neural activity of the considered networks\cite{golomb1993dynamics,xu2021diversity}, where:

\begin{align}\label{eq:definition_index}
\chi^2=\dfrac{N\sigma_{V(t)}^2}{\sum_{i=1}^N\sigma_{V_{i}(t)}^2}
\end{align}
Here $V(t) = \frac{1}{N}\sum_{i=1}^NV_{i}(t) $ denotes population average of the membrane potential $V_{i}(t)$.  $\sigma_{V(t)}$ and $\sigma_{V_{i}(t)}$ are standard deviation of $V(t)$ and the membrane potential traces $V_{i}(t)$ over time. $\chi$ is $1$ when all the neurons have the same trajectory and $0$ for an incoherent state when the fluctuations of $V(t)$ are $0$.

To further investigate the global dynamical behavior of the neural networks, we   further introduce the order parameter\cite{kuramoto2003chemical,bertolotti2017synchronization}, $R$, together with  its the variance in time, metastability\cite{shanahan2010metastable,xu2018synchronization,xu2021diversity}, $Met$, given by:

\begin{align} 
R &= \left\langle \left| \frac{1}{N_E} \sum_{k=1}^{N_E} e^{2\pi i[(t-t_k^n)/(t_k^{n+1}-t_k^n)]}  \right| \right\rangle   \label{eq:definition_R}  \\  
Met & = \frac{1}{T} \sum \limits_{t\leq\Delta t}\left(\phi_c(t)-R\right)^2    \label{eq:metastability}
\end{align} 

The angle brackets in \hyperref[eq:definition_R]{Eq.(\ref{eq:definition_R})} is the temporal average value of that quantity and $2\pi[(t-t_k^n)/(t_k^{n+1}-t_k^n)]$ is the phase of each neurons in  the excitatory  population.  $t_k^n$ is coming time of the $n$th spikes of neurons $k$ (and thus $t_k^{n+1}$ is time of the following $(n+1)$th spikes). The closer to $R=0 (1)$ becomes, the more asynchronous (synchronous) the dynamics is. The global metastability\cite{shanahan2010metastable}, $Met$ (\hyperref[eq:metastability]{Eq.(\ref{eq:metastability})}) is used here to quantify  metastability and chimera-likeness of the observed dynamics. Metastability is $0$ if the system is either completely synchronized or completely desynchronized -- a high value is present only when periods of coherence alternate with periods of incoherence.

  We next introduce SPIKE-Synchronization (SPIKE-Syn)   for quantifying similarity in terms of the fraction of coincidences between two spike trains\cite{kreuz2015spiky,mulansky2016pyspike}. A coincidence indicator $C_{i}^{1,2}$ for SPIKE-Syn is described for every spike of the two spike trains $S_{i}^{1,2}$. The $C_{i}=1$ if the spikes at $t_i$ is part of a coincidence and $C_i = 0$ if not. The value of the coindidence indicator is then given by:
\begin{equation}
C_i^1=\left\{
\begin{aligned}
1 & \quad \text{if} \min_{j}(|t_i^1-t_j^2|) < \tau_{ij}^{1,2}\\
0 &  \quad \text{otherwise}
\end{aligned}
\right.
\end{equation}
 Where $\tau_{ij}^{1,2}= \frac{1}{2}\min\{\nu_{i}^1,\nu_{i-1}^1,\nu_{j}^2,\nu_{j-1}^2\}$ is an adaptive coincidence window according to the local firing rate.  the interspike intervals are given as $\nu_i^{1,2}=t_{i+1}^{1,2}-t_{i}^{1,2}$.  The coincidence indicator for the second spike train $C_i^2$ is computed as the same way. The spike-synchronization profile is then given by the discrete function in terms of the pooled coincidence indicators $\{C_k\} = \{C_i^1\} \cup \{C_i^2\}$ and spike times $\{t_k^{'}\} = \{t_i^1\} \cup \{t_i^2\}$. The pairwise time spike synchronization values (spike$\_$sync$\_$matrix) $s_c$,quantifies the fraction of all spikes in the two spike trains  that are coincident, given by:

\begin{align} 
s_c= \frac{1}{M}\sum_{k=1}^{M} C_{k}
\end{align} \label{eq:sync}
where $M = M_1+M_2$ denoting the total number of spikes in the pooled spike train.  $s_c=0$ for spike trains without any coincidences and $s_c=1$ if and only if the two spike trains consist only of pairs of coincident spikes. The generalizing coincidence indicators is defined as before: $ C_i^{n,m} $ is 1  if  $\min_{j}(|t_i^n-t_j^m|) < \tau_{ij}^{n,m}$ and 0 if not, where $\tau_{ij}^{nm}$ is similar as above. A normalized coincidence counter
for each spike of every spike train is given as $C_i^n = \frac{1}{N-1}\sum_{m\neq n}C_i^{n,m}$ obtained by averaging over all $N-1$ bivariate coincidence indicators involving the spike train $n$. Therefore , the multivariate SPIKE-synchronization  is described by:

\begin{align} 
S_C= \frac{1}{M}\sum_{k=1}^{M} C_{k}\label{eq:syncspike}
\end{align} 
where $M= \sum_n^m M_n$ again denotes the overall number of spikes.  The interpretation is very intuitive: SPIKE-Synchronization quantifies the overall fraction of coincidences. It is zero if and only if the spike trains do not contain any coincidences, and reaches one if and only if each spike in every spike train has one matching spike in all the other spike trains.

\begin{figure}[htp]	
\includegraphics[width=\linewidth]{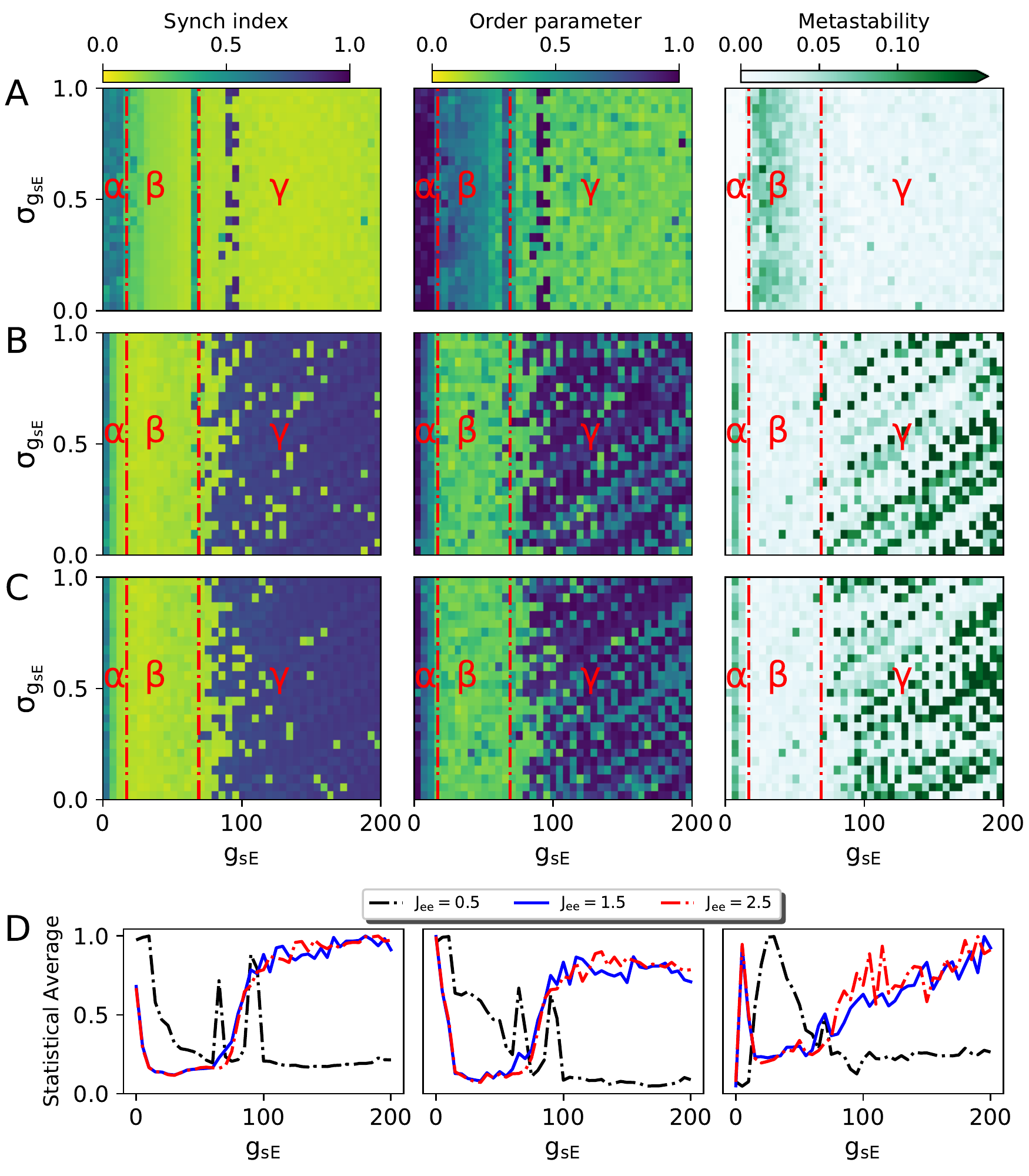}	  
	\caption{The effect of chemcial synaptic noise among excitatory neurons on coexisting firing patterns in $g_{sE}/\sigma_{g_{sE}}$ parameter space.   The variables $g_{sE}$, $\sigma_{g_{sE}}$ are the means and standard deviation  of  Gaussian distribution uesd  here for peak synaptic conductances. In presence of chemical synaptic noise among excitatory neurons can produce significant  emergence of neural network state, that are:  \textbf{A}.($\alpha$). synchronous firing pattern,($\beta$). metastable  state, ($\gamma$).asynchronous firing state;  \textbf{B-C}.($\alpha$).coexistence of synchronous  and asynchronous firing pattern, ($\beta$).asynchronous firing state, ($\gamma$).coexistence of metastable and synchrony state.   \textbf{D.} Statistical average of synchrony index, order parameter and metastability (left to right) over all the $\sigma_{g_{sE}}\in(0,1]$ against $g_{sE}\in(0,200]$. The excitatory  synaptic weights of E-population  for each row  are  \textbf{A.} $J_{EE}=0.5$, \textbf{B.} $J_{EE}=1.5$, \textbf{C.} $J_{EE}=2.5$.}
	\label{fig:effectnoise}
\end{figure}

\section{Influence of synaptic noise among excitatory neurons in coexisting firing patterns}

Networks with mixed excitatory/inhbitory (E/I) neurons, displaying various firing patterns depending on the presence of gap junction  in the E/I balance network, has been  well studied in our previous work\cite{xu2021diversity}. In this work, we will investigate the  contribution  of chemical synaptic noise on  coexisting firing patterns of a hybrid synaptic interacting  balanced neuronal network. Fig.\hyperref[fig:effectnoise]{\ref{fig:effectnoise}A} shows  collective behaviour of coexisting firing patterns resulting from the simulation of a network consisting on 1,000 excitatory and 250 inhibitory neurons, in a full region of the $g_{sE} /\sigma_{g_{s}E}$ parameter space. We characterized these firing regimes by calculating the synchronization index $\chi$ (in Eq.\ref{eq:definition_index}) for  voltage traces, and the Kuramoto order parameter $R$ for phase synchronization based on spike firing (see Eq.\ref{eq:definition_R}). We also  introduce the metastability $Met$  \hyperref[eq:metastability]{Eq.(\ref{eq:metastability})} to quantify the metastability and coexisting firing pattern  of the observed dynamics within the excitatory population. The results from a balanced neuronal network with weaker excitatory connection, when $J_{EE}=0.5$ as a example, is shown in Fig.\hyperref[fig:effectnoise]{\ref{fig:effectnoise}A}.  Fig.\hyperref[fig:effectnoise]{\ref{fig:effectnoise}A} shows three significant but different phases of collective network states that are synchronization state,  time-varying coexisting firing state (or multistable state) and asynchronous state, characterizing behavior of the neuronal networks by means of the synchrony and metastability indexes. The general synchronization index $\chi$ is highly dependent on the  means value $g_{sE}$ of Gaussian distribution, displaying three distinct collective behaviour with color regions  shown as $\alpha,\beta,\gamma$. However, for a given value of $g_{sE}$, neural firing and synchronization modes of these three regions are robust against different values of noise variance, $\sigma_{g_{sE}}$.  This observation suggests  that the means $g_{sE}$ of Gaussian distribution as source of synaptic noise make a greater contribution than standard deviation $\sigma_{g_{sE}}$  to neural firing patterns.   The synchronization of spikes quantified  by  order parameter $R$  also shows a similar pattern although these three firing regions appear less homogeneous than in the case of the general index $\chi$. More importantly, the metastability index shows this, with a metastable region separating the  synchrony region $\alpha$ and  asynchronous region $\gamma$. As predicted by the measurement of statistical methods mentioned above, an increase in mean $g_{sE}$ of peak synaptic conductances causes the appearance of a region $\beta$ with higher metastability  as a  transition from synchronous  to  asynchronous firing. 
     
     The effect of increasing synaptic weights among excitatory neurons on firing pattern  is exhibited  in Fig.\hyperref[fig:effectnoise]{\ref{fig:effectnoise}}B-C).There is a dramatic changes for $\gamma$ regimes in the synchrony measures although this effect is not evenly distributed across the whole parameter space.Region $\gamma$ shows a high increase in synchrony, displaying as speckle pattern of parameter space. As hinted by this speckle pattern, higher values of this pattern, approximately equal to 1, are  often characterized by coherent state of activity that can not find the metstable or asynchronous behaviour;  whereas the rest region in the speckle pattern is an intermediate synchrony region, indicating metstable state that is  characterized by high metastability. In the other words,  speckle pattern shown in $\gamma$   indicates the coexistence of synchronous firing and metstable state with slightly varying systems parameters, which is namely parameter-varying coexistence.  The  more important findings  shown  in speckle pattern (Fig.\hyperref[fig:effectnoise]{\ref{fig:effectnoise}}B,C) is  that both means $g_{sE}$ and $\sigma_{g_{sE}}$ of Gaussian distribution can play a great role on the neural firing pattern, which is different in case of weaker excitatory weights plotted in Fig.\hyperref[fig:effectnoise]{\ref{fig:effectnoise}}A. On the contrary, region $\beta$ in \hyperref[fig:effectnoise]{Fig.\ref{fig:effectnoise}}B,C displays a decrease of synchronization index and order parameter, as a result of increasing the excitatory synaptic weights. The results obtained from metastability (Fig.\hyperref[fig:effectnoise]{\ref{fig:effectnoise}}B,C)  imply that region $\beta$ goes to an asynchronous firing state instead of metstable state in Fig.\hyperref[fig:effectnoise]{\ref{fig:effectnoise}}A with slight enhencing $J_{EE}$. Region $\alpha$ shows  coexistence of synchronous and metastable state (Fig.\hyperref[fig:effectnoise]{\ref{fig:effectnoise}}B,C) instead of absolutely synchronization, characterized statistical methods mentioned. Finally, in order to further understand the effect of excitatory synaptic weights on neural firing mode, Fig.\hyperref[fig:effectnoise]{\ref{fig:effectnoise}}D plots statistical average of synchrony index, order parameter and metastability over $\sigma_{g_{sE}}$ with growth of $g_{sE}$. As predicted, the results obtained from $\alpha,\beta,\gamma$ regions  are robust with enhancing excitatory connection weights shown in Figs.\hyperref[fig:effectnoise]{\ref{fig:effectnoise}}B,C. However, when  E-population  with weak coupling(in Fig.\hyperref[fig:effectnoise]{\ref{fig:effectnoise}}A), these regions exhibit their opposite impact on neural responses, clarifying that the repetitive firing properties of  E-population is greatly affected by excitatory synaptic weights. In all, exploring the $g_{sE}/\sigma_{g_{sE}}$ parameter space  under suitable excitatory synaptic weights is beneficial for the emergence of novelty firing patterns, such as parameter-varying coexistence of metastable  and synchrony state that can not be observed with weaker excitatory coupling.

    In the following,  we will examine in more detail the coexistence of several possible firing patterns and their transitions between these firing regimes caused by  increasing  the means of Gaussian distribution. To do this, we will focus on two values of excitatory strength: $J_{EE}=0.5$ and $J_{EE}=1.5$, roughly sweeping the areas of a widely  range $g_{sE}\in(0,200]$.

\section{Repetitive firing patterns and transitions between different firing regimes}
\begin{figure}[htp]	
\includegraphics[width=\linewidth]{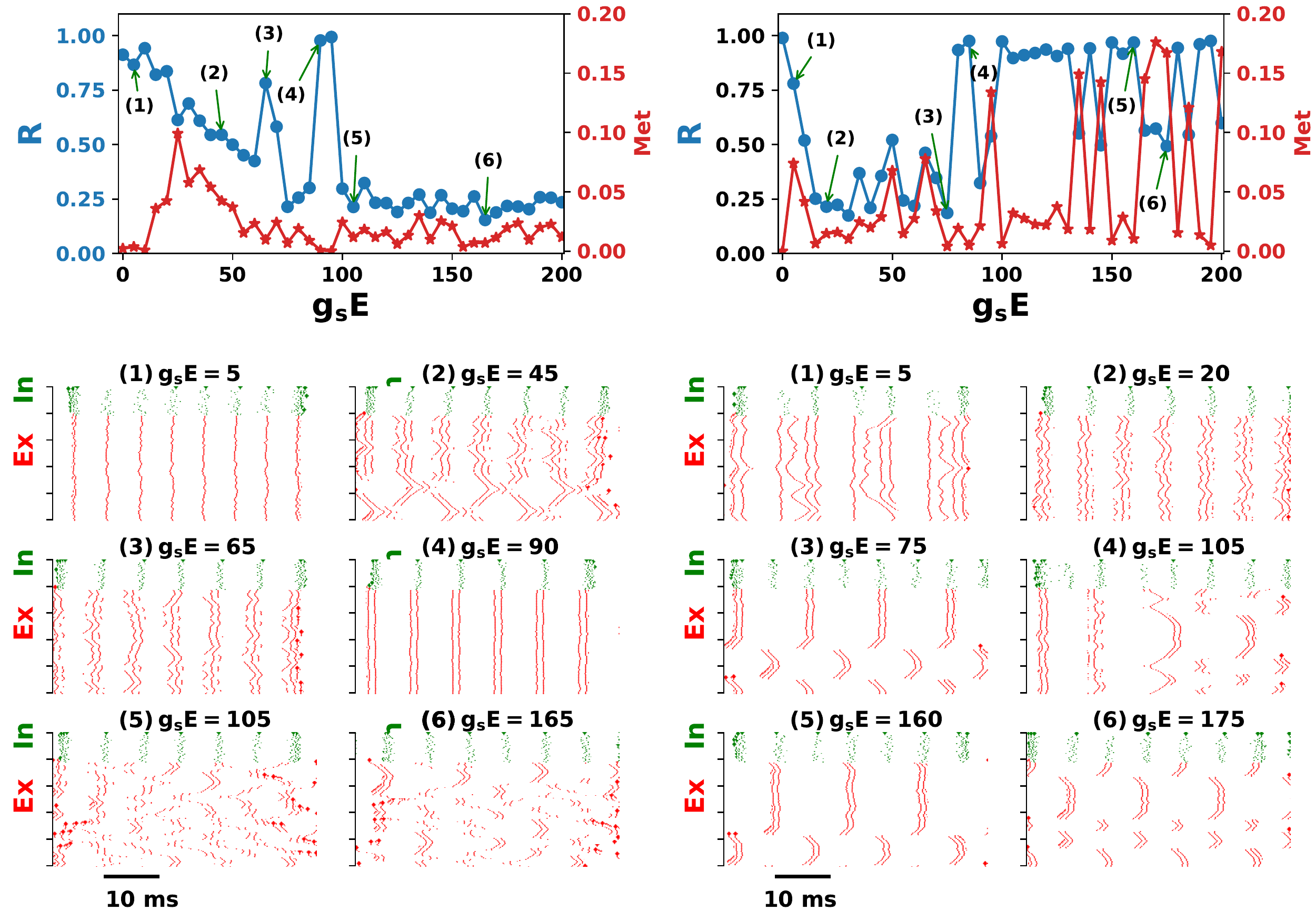}	  
	\caption{Significant emergence of  neural network state and transitions with enhencing synaptic noise under different level of excitatory coupling, where $J_{EE} = 0.5$ (\textbf{A,C}) and $J_{EE} = 1.5$ (\textbf{B,D}).   Red star and blue circle lines represent  evolution of order parameters and metastability for excitatory population. \textbf{C} and \textbf{D}  show a detail examples of typical spatiotemporal firing patterns that corresponding to letters shown in (\textbf{A}) and (\textbf{B}) in each dynamics regimes as sweeping parameter $g_{sE}$.  (a,d): Synchronous states;  (c,g,h) Generalized synchrony  states;   (e,f): asynchronous states; (b,i,j,k,l) Metastable state or chimera-likeness. }
	\label{fig:firingmode}
\end{figure}

    We then investigated various firing patterns at different excitatory level, to characterize the firing patterns and their transitions between these firing state that represented by $\alpha,\beta,\gamma$ shown in Fig.\hyperref[fig:effectnoise]{\ref{fig:effectnoise}}. Fig.\hyperref[fig:firingmode]{\ref{fig:firingmode}} shows the evolution of synchrony of order paramenter for spikes and metastability as $g_{sE}$ is swept from 0 to 200, under weak (Fig.\hyperref[fig:firingmode]{\ref{fig:firingmode}}A,C) and  a slightly stronger(Fig.\hyperref[fig:firingmode]{\ref{fig:firingmode}}B,D)  excitatory coupling.

    With weak (Fig.\hyperref[fig:firingmode]{\ref{fig:firingmode}}A,C) and  a little stronger excitatory synaptic coupling (Fig.\hyperref[fig:firingmode]{\ref{fig:firingmode}}B,D),  the E-population of balance network usually can go through different phases of network state as variation of $g_{sE}$. The letters  a-l in Fig.\hyperref[fig:firingmode]{\ref{fig:firingmode}}C,D  represent some examples of various firing pattern for each dynamical regiems  that shown in  \hyperref[fig:effectnoise]{Fig.\ref{fig:effectnoise}}.   The activities of synchronous, metstable and asynchronous states have been clearly observed looking at variation through $R$ and $Met$.  These significant  dynamical regimes  as well as corresponding transitions can be defined based on the different values of $Met$ . The finer sweep of the means of synaptic noise now allows to observe that the transition between the one-spike synchronous firing pattern (a) and the two-spike synchronous pattern (d) occurs with a  little more disordered patterns (b,c).  These  disordered firing patterns  are characterized by high metastability, exhibiting time-varying  coexistence of unstable and transient traveling waves. It is worth noting that  two-spike synchronous pattern(d) appear here also as a transition to the asychronous firing oscillatory regime(e,f), characterized by low order paramter and metastability when $g_{sE}>95$. In contrast, we repeated the same exploration at a higher level of excitatory strength when $J_{EE}=1.5$ in \hyperref[fig:firingmode]{Fig.\ref{fig:firingmode}B,D}.  The resulting  varying values of $R$ and $Met$ against $g_{sE}$ , as well as rastor plots, show that the activity of parameter-vaying coexisting firing is  incoherent,  that are coexistence of  generalized synchrony  states (g,h) and metstable state (i,j,k,l) at different values of $g_{sE}$.  Moreover, we observe  that  metstable state  has been   performed  as format  of  coexistence  or  alternation  of interest between synchronous state and traveling wave under suitable parameter values of $g_{sE}$.  In all,  the main findings  in \hyperref[fig:firingmode]{Fig.\ref{fig:firingmode}}  provide supporting evidence that synaptic  noise intensity  and  excitatory level  are  significant system parameters  which play a major role in determining the emergence of the firing patterns, including traveling wave, synchrony state with one-spike (or two-spike),  a peculiar  coexisting firing patterns  (chimera-likeness behavior). More importantly, neural firing transitions between dynamical states  also depend on strength of synaptic noise  among individual neurons. The  more detailed  influence of the  whole   $g_{sE}/J_{EE}$  parameters  space on neuronal firing patterns  will be systematically studied  in section \ref{excitatorycoupling}.

\begin{figure}[htp]	
\includegraphics[width=\linewidth]{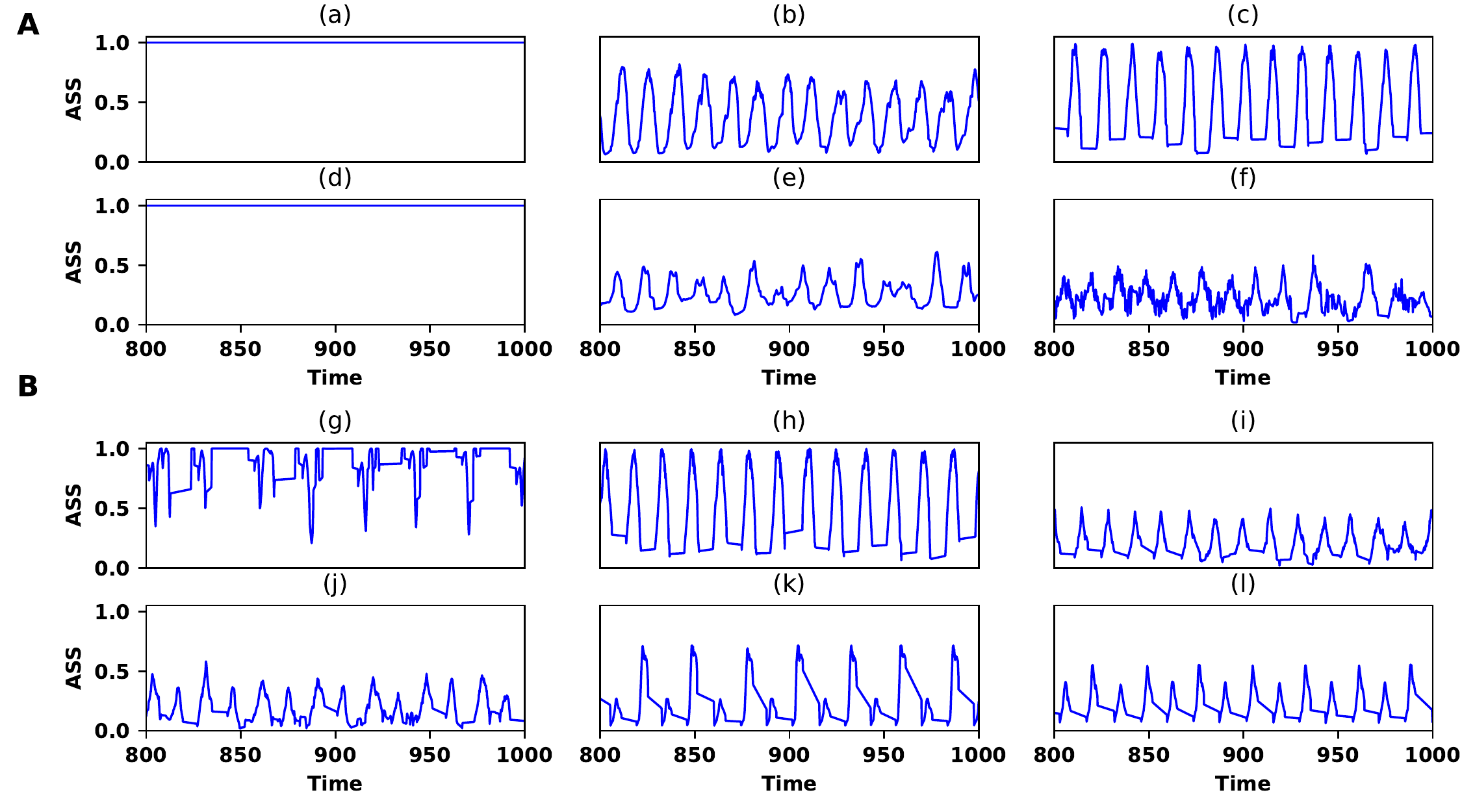}	  
	\caption{Averaged SPIKE-Sync profile (ASS) for a fixed time window,  quantifying   similarity  and  periodicity (or quasi-periodic )  of neuronal firing state in time. Subplots \textbf{A}  and  \textbf{B} respectively corresponding to  the  raster plots of spike train that  plotted in \hyperref[fig:firingmode]{Fig.\ref{fig:firingmode}C,D }. The letters $a-l$ denote   the  examples of firing rate  in  \hyperref[fig:firingmode]{Fig.\ref{fig:firingmode} }.} 
	\label{fig:spikesync}
\end{figure}

       So far we have focused  on conditions of  arising  coexisting firing patterns  and  statistical methods of measuring the degree of synchronization in E-population. We then computed  SPIKE-Sync profile  in Eq.\ref{eq:syncspike}  of  the excitatory spike strains, as well as  average SPIKE-Sync  shown as function of time,   to  quantify   similarity  and  periodicity  of neuronal firing state.  \hyperref[fig:spikesync]{Fig.\ref{fig:spikesync}}  plots  averaged SPIKE-Sync profile (ASS) for a fixed time window to  quantify  similarity  and  periodicity  of neuronal firing state, that that are shown in  \hyperref[fig:firingmode]{Fig.\ref{fig:firingmode}C,D } .  With weak coupling (\hyperref[fig:spikesync]{Fig.\ref{fig:spikesync}A}), the balanced neural network  exhibited higher correlated spike trains (\hyperref[fig:spikesync]{Fig.\ref{fig:spikesync}(a),(d)}), which is steady-state synchronization characterized by $ASS = 1$,  in contrast to spike trains in the asynchronous state (\hyperref[fig:spikesync]{Fig.\ref{fig:spikesync}(e),(f)}).  As in the asynchronous state,  irregular activity with lower value  of  averaged SPIKE-Sync profile  is observed, denoting   lower and weaker correlated spike trains along time.  More importantly,  varying  $g_{sE}$ demonstrates  that the  balanced network also can  exhibit  similarity,  and 1-periodic (or quasi-periodic ) acitivity  of great  interest (\hyperref[fig:spikesync]{Fig.\ref{fig:spikesync}(b),(c)}), implying again emergence and multistability of time-periodic states in a excitatory population of noise-driving and hybrid synaptic interacting networks (see \hyperref[fig:firingmode]{Fig.\ref{fig:firingmode}C }). 
       
     As predicted  in \hyperref[fig:spikesync]{Fig.\ref{fig:spikesync}B} at a higher level of excitatory coupling, the big difference is that  only weaker correlated  (or  uncorrelated) spike trains have been observed from ASS time series with increasing $g_{sE}$. It was  surprisely found that the repetitive  behaviors   with  2-periods  are shown  in \hyperref[fig:spikesync]{Fig.\ref{fig:spikesync}(i-l)}  by the spike train similarity measures  of that $ASS <1$, suggesting  the  emergence of   periodic  firing patterns  accompanying with  subthreshold  oscillations  of some  neurons  in a  excitatory population (compares to i-l  in  \hyperref[fig:firingmode]{Fig.\ref{fig:firingmode}D} ).   Secondly, quasi-periodic  firing patterns   also has been observed  in  \hyperref[fig:spikesync]{Fig.\ref{fig:spikesync}(h)}.  Otherwise, alternative time series  of  ASS between synchronous  and asynchronous state has been shown in \hyperref[fig:spikesync]{Fig.\ref{fig:spikesync}(g)}.

\begin{figure}[htp]	
\includegraphics[width=\linewidth]{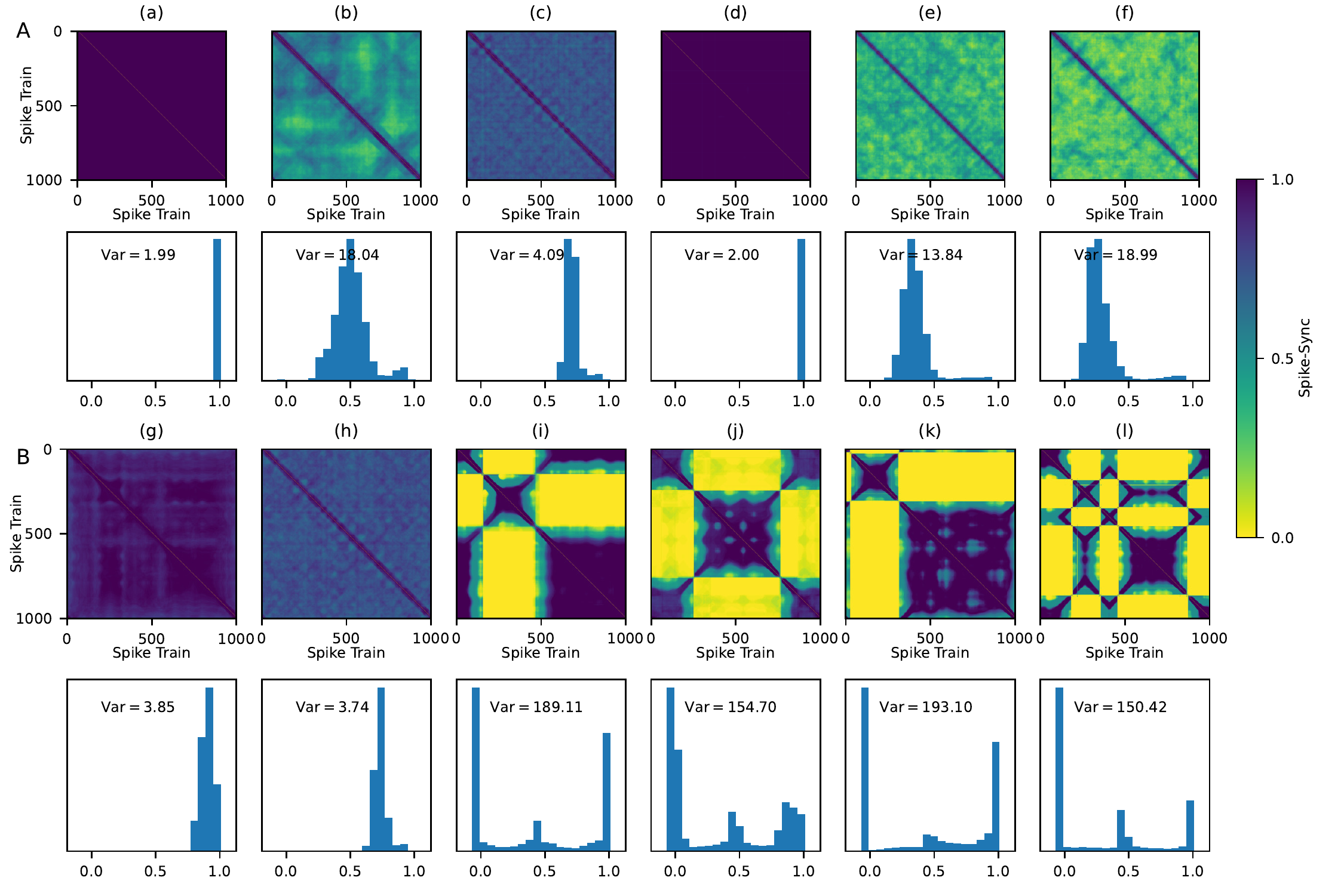}	  
	\caption{Dissimilarity matrices obtained by  pairwise  spike-sync of the  spike trains of  excitatory population.  \textbf{A} and \textbf{B}, top: SPIKE-sync  matrices obtained at  different values of  synaptic noise  strength  $g_{sE}$ in   balanced neuronal  networks of  hybrid synaptic connections. Below each  matrix, an histogram of the spike-syn values is shown.   Values of  Variance, Var ( = Variance*1000),  are shown inside each  histogram plot.   \textbf{A}. $J_{EE}= 0.5$ ; \textbf{B}. $J_{EE} =1.5$ . } 
	\label{fig:SPIKESyncmatrix}
\end{figure}

     To demonstrate  groups  (or clusters)    as well as   transitions  of  various  firing pattern that already observed,  dissimilarity matrices obtained by  pairwise  spike-sync in Eq.\ref{eq:syncspike} have been  introduced.   Dissimilarity matrices  (Fig.\hyperref[fig:spikesync]{\ref{fig:SPIKESyncmatrix}}) show distinctive patterns for the unsynchronized and synchronized situations  at  two different level  of  excitatory strength.In the first case of  higher correlated state, complete (Fig.\hyperref[fig:spikesync]{\ref{fig:SPIKESyncmatrix}(a),(d)})  and moderate  synchronous state  (Fig.\hyperref[fig:spikesync]{\ref{fig:SPIKESyncmatrix}(c),(g),(h)})   have been clearly  detected  depending on values outside the diagonal and their variances.  On the other hand, for  complete  synchronous state, all the values in the dissimilarity matrix are equal to 1, meaning that the synchronization is the same and maintained through all the simulation; for  moderate  synchronous state,  dissimilarity  matrices show a mixture of values between 0.5 and 1,  that evidence  a lower degree of  synchrony firing patterns.  In the second case of  uncorrelated state,  asychronous firing state and multistable state  have already been observed.    For weaker excitatory connections  (shown in Fig.\hyperref[fig:spikesync]{\ref{fig:SPIKESyncmatrix}A}),   dissimilarity  matrices with smaller $g_{sE}$ show  intermediate values around 0.5, that imply  a coexisting firing patterns of  multistable regime  as a transition between complete and moderate  synchronous state  shown in Fig.\hyperref[fig:spikesync]{\ref{fig:SPIKESyncmatrix}(b)}.   Moreover, at the large $g_{sE}$, values  in dissimilarity  matrices are almostly less than 0.5 and the mean  of SPIKE-Syn is around 0.25 , that are asychronous firing state shown in Fig.\hyperref[fig:spikesync]{\ref{fig:SPIKESyncmatrix}(e),(f)}. However,  for a little stronger excitatory connections (see Fig.\hyperref[fig:spikesync]{\ref{fig:SPIKESyncmatrix}B}), dissimilarity  matrices display some clusters of  the values between 0 and 1  at the large $g_{sE}$ , with noticeable regular ‘patches’ shown in Fig.\hyperref[fig:spikesync]{\ref{fig:SPIKESyncmatrix}(i) -(l)}  that evidence the maintenance of some synchronization patterns. The histograms of SPIKE-syn values (shown in  Fig.\hyperref[fig:spikesync]{\ref{fig:SPIKESyncmatrix}} below each dissimilarity matrice) are also useful in detecting the three situations described.

\begin{figure}[htp]	
\includegraphics[width=\linewidth]{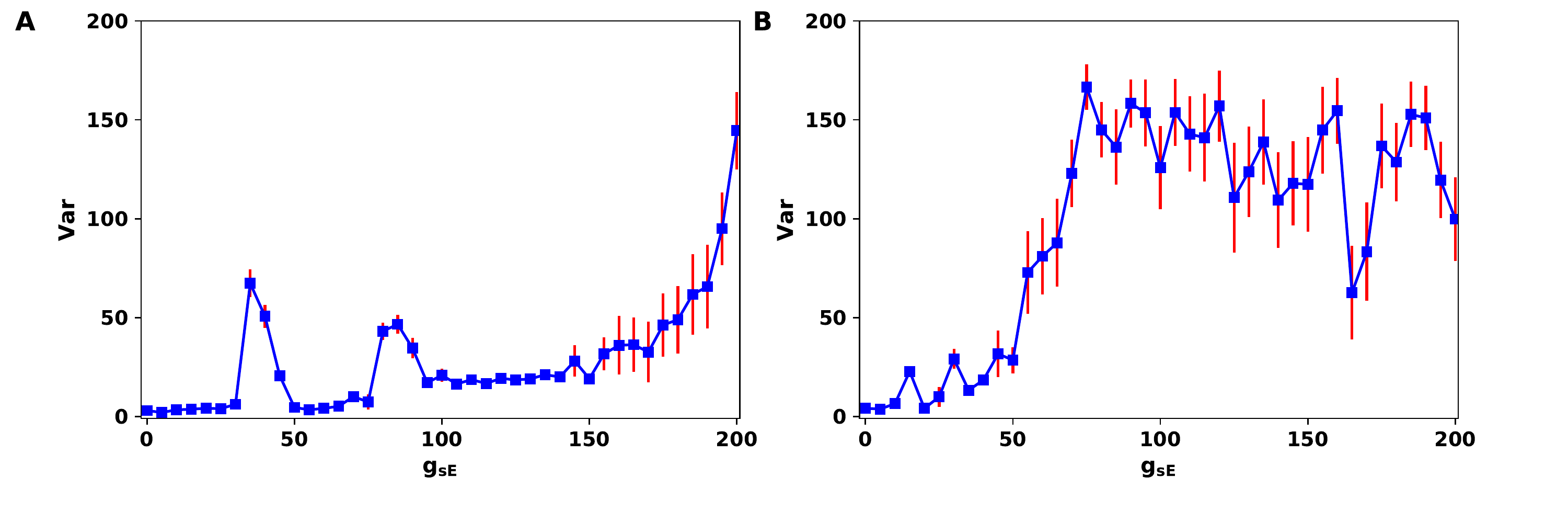}	  
	\caption{Variance of the SPIKE-Sync  values (the same values plotted in the histograms) plotted against $g_{sE}$, where Var shown as ylabel is defined as: Var =Variance of SPIKE-Sync  values*1000.  Average (±SEM) of 10 simulations with different random seeds for the small-world connectivity and parameters.\textbf{A}. $J_{EE}= 0.5$; \textbf{B}. $J_{EE} =1.5$.} 
	\label{Fig6Vari}
\end{figure} 
    
      As a rough measure of multistability, we took the variance of the SPIKE-Syn values (outside the diagonal)  and plotted them against the strengths of  synaptic noise, averaging several simulations  with different seed for the random connectivity matrix(Fig.\hyperref[Fig6Vari]{\ref{Fig6Vari}}). Var is $0$ if the system is either completely synchronized or completely desynchronized  and a high value is present only when periods of coherence alternate with periods of incoherence. At weak excitatory coupling (Fig.\hyperref[Fig7Vari]{\ref{Fig6Vari}}A), it is clear that  the neuronal networks can produce time-varying coexisting firing patterns of that Var $>$0, which is investigated and shown  in Fig.\hyperref[fig:firingmode]{\ref{fig:firingmode}} ,  for some suitable values of  $g_{sE}$. Besides, transitions between different  neuronal firing  patterns  are clearly  exhibited with varying $g_{sE}$. With high level of excitatory coupling (Fig.\hyperref[Fig6Vari]{\ref{Fig6Vari}}B),  both time-varying and parameter-varying coexisting firing patterns  have  been  clearly observed  in  a  wider $g_{sE}$ range,  particular when  $g_{sE} \geq 50$ of  Var $>>$0. Overall, the SPIKE-Syn values with signatures of  coexisting firing activies  confirms again that  multistable behaviour are more easily obtained when the networks are at different level of excitatory coupling, while parameter-varying coexistence of synchrony and various ripple events are more easily observed  when  the networks are at high level excitatory coupling.

\section{Effect of excitatory coupling  strengths  in the generation of coexisting  dynamical  regimes }
\label{excitatorycoupling}
\begin{figure}[htp]	
\includegraphics[width=\linewidth]{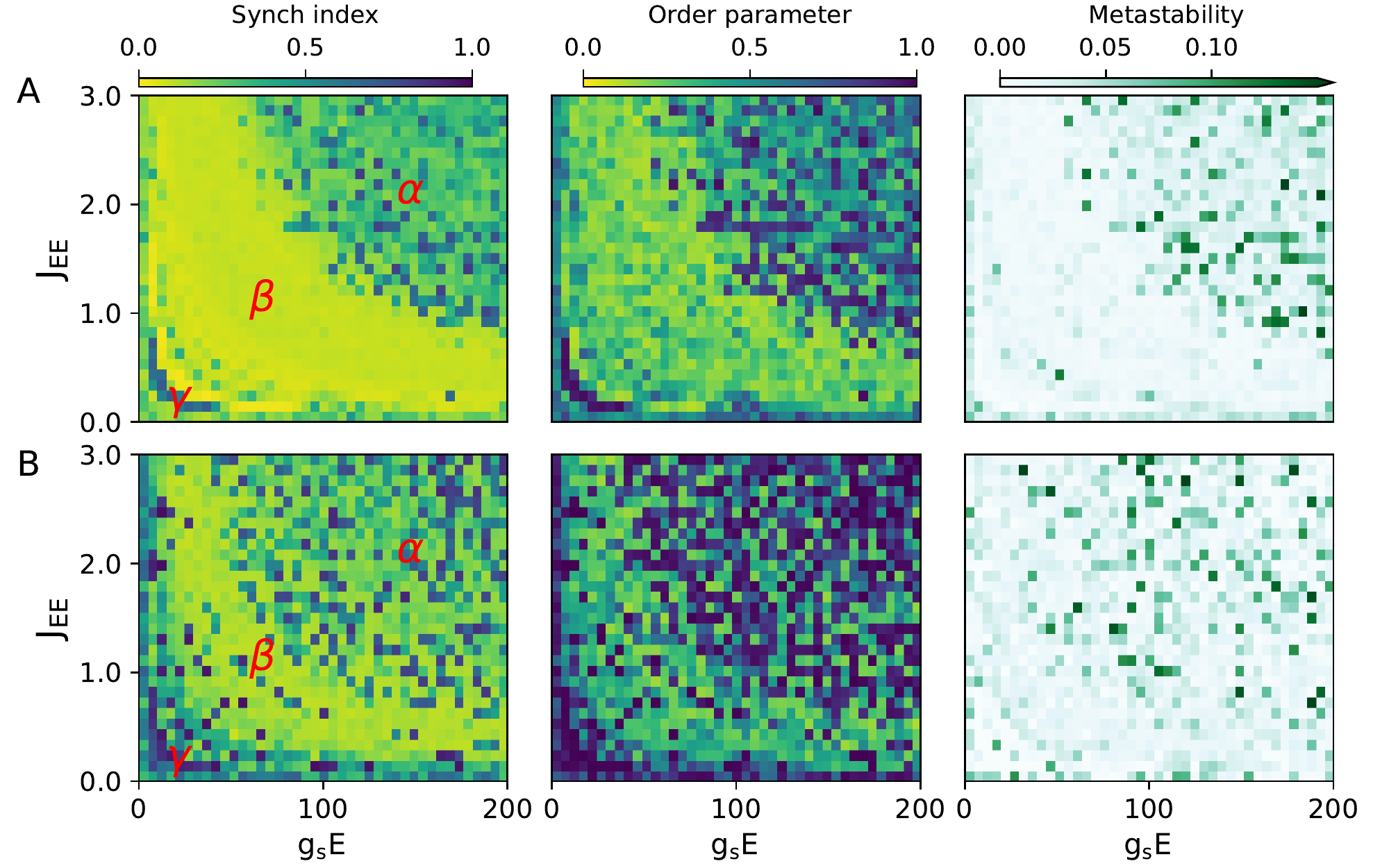}	  
	\caption{The influence of  excitatory  synaptic  weights   on  coexisting firing patterns in $g_{sE}/{J_{EE}}$ parameter space at two different level of electrical coupling.  Regions $\alpha$ and $\gamma$ denote coexisting firing patterns, showing the similar dynamical behaviour  that observed in Fig.\hyperref[fig:effectnoise]{\ref{fig:effectnoise}}.Region $\beta$ exhibits asychronous firing state with lower values ($\approx 0$)  of statistical parameters.  Weights of electrical synaptic coupling are:  \textbf{A.}$J_{gap}=0.1$ ;  \textbf{B.}$J_{gap}=1$.} 
	\label{fig:ExcSCI}
\end{figure} 

To present a broader perspective on variety of firing patterns  that  affected by combination of  chemical synaptic noise  and strengths of excitatory connections,  we explore the  synchronized and asynchronized states  as well as coexisting firing patterns  in  $g_{sE}/{J_{EE}}$  parameter space.  Fig.\hyperref[fig:ExcSCI]{\ref{fig:ExcSCI}A,B}   shows  the collective behavior  of excitatory  neurons  at two different level of  electrical coupling, respectively. With weak electrical synaptic coupling  of that $J_{gap} = 0.1$, it is clear that  both  asynchronized states (green- yellow region, $\beta$)  and  'speckle' pattern of coexisting firing state (regions $\alpha$ and $\gamma$)  have been observed with sweeping $g_{sE}/J_{EE}$  shown in  Fig.\hyperref[fig:ExcSCI]{\ref{fig:ExcSCI}A}(left). Moreover, coexisting firing patterns  of interest induced by synaptic noise  separates into two distributaries (regions $\alpha$ and $\gamma$) through asynchronized firing state of   region  $\beta$. In other words, the asynchronized firing regimes emerges as a transition  among  coexisting firing state  shown in Fig.\hyperref[fig:ExcSCI]{\ref{fig:ExcSCI}A}(left).The results of these finding also has been evidenced by a statistical methods of order parameter and metstability shown in the  Fig.\hyperref[fig:ExcSCI]{\ref{fig:ExcSCI}A}(middle and right). However, in the case of stronge electrical synaptic  coupling when $J_{gap} = 1$, Fig.\hyperref[fig:ExcSCI]{\ref{fig:ExcSCI}B} exhibits that asychronous firing regimes have been gradually displaced by coexisting firing patterns,that will cover the overall  $g_{sE}/J_{EE}$ parameter space eventually with further increasing electrical synaptic weights. At end,  our findings  imply that  synaptic noise, excitatory synaptic weights  as well as electrical coupling strengths   make a  significant contribution to the generation of  coexisting firing patterns in the hybrid synaptic interacting  balanced neuronal network.

In summery, we investigated  multistability or coexistence of  neural firing pattens  that driven by chemical synaptic noise in a  balanced network of excitatory neurons connected by  hybrid synapses. Using statistical measurement of  synch-index, order parameter , metastability and SPIKE-syn,  we found that  strengths of synaptic noise and excitatory weights  make a greater contribution than variance of synaptic noise to the coexistence of firing states  with slight modification parameters.  Specifically, the  coexisting  firing patterns in a hydrid coupling neural networks can  emerge  in  two ways, including time-varying and parameter-varying coexisting firing patterns. The first type of coexisting  firing patterns  in neural systems is typically metstable state as format of  the time-varying coexistence, namely coexistence or alternation of synchronous state and traveling wave.   The presence
of  this coexistence firing patterns has been observed under the fixed  combination of  synaptic noise  strengths and excitatory connection weights. The second type of multistability in neural networks is parameter-varying coexistence.  The  parameter-varying coexistence is  accompanied by coexistence of coherence(synchrony state) and incoherence  firing state(metastable or asynchronous firing pattern) with slightly varying control parameters.

\section{summery and discussion}
  The phenomenon of multistability has been found in various classes of dynamical systems such as delayed feedback systems,weakly coupled systems, excited systems, and stochastic systems\cite{feudel2008complex,stankovski2017coupling}.The appearance of multistability depends on many factors such as the strength of dissipation,the nature and strength of coupling, the value of the time delay, amplitude and frequency of the parameter perturbation,and noise intensity\cite{pisarchik2014control,kim1997multistability}. Experiments as well as theoretical models have revealed different routes to multistability in the different system classes(see review\cite{pisarchik2014control}).In previous works\cite{orio2018chaos,xu2021diversity} and in the present one, we try to unveil
the impact of synaptic connections as well as the node dynamics to the various firing activities and the multistable dynamics found in a hybrid synaptic interacting balanced neuronal network.

  A wealth of evidence indicates  abundant electrical synapses interlinking excitatory neurons.Electrical synapses occur between excitatory glutamatergic inferior olivary cells \cite{llinas1974electrotonic}, glutamatergic excitatory trigeminal mesencephalic primary afferent neurons\cite{hinrichsen1970coupling}, and others(reviewed in\cite{nagy2018electrical}).Nevertheless,theoretical understanding of hybrid synaptic connections
in diverse dynamical states of neural networks for self-organization and robustness still has not been fully studied. In order to reveal the underlying roles of this mixed synaptic connection on various firing patterns, we  have already presented a model of neural network  including chemical excitatory and electrical synaptic coupling for excitatory population in our previous work\cite{xu2021diversity}. In this balanced neural networks, we  found that 
the emergence of various firing ripple events  has been observed by considering the variation of chemical synaptic inhibition and  network densities. Secondly,  we also can see that the excitatory population has a tendency to synchronization as the weights of electrical synaptic coupling among excitatory cells are increased. Moreover, the existence of
this mixed synaptic connections in excitatory population  can cause various firing patterns of interest by slightly changing the chemical synaptic weights. As we know, observation of
neural noise has been explored in multiple experimental studies from the perception of sensory signals to the generation of motor responses\cite{faisal2008noise,mcdonnell2011benefits}. However,  effect  of synaptic noise  on  genertion of coexisting firing patterns in a hybrid  connected balanced networks has not  well been considered in that  study.

It is generally accepted that coexisting firing patterns induced by noise or chaos can arise in networks purely connected by electrical or chemical synapses together with synaptic weights and seems to depend on other factors such as time delay and networks topology\cite{baptista2010combined,sporns2010networks}. Computational models show promise to identify some of underlying mechanisms to reproduce coexisting firing patterns. Early studies have reported  that the inclusion of noise  was regarded as an ad hoc mechanism required to produce interesting dynamics\cite{deco2012ongoing,golos2015multistability,hansen2015functional,deco2017dynamics}. A common feature of the noise-driven multistable models is the existence of a subcritical Hopf bifurcation to generate multiple attractors. Then, noise produces the switching between them. In other words,  for deterministic models attracted to stable or periodic solutions, noise is fundamental to avoid that dynamics become stuck in a state of equilibrium. Therefore, the ad hoc introduction of noise in a dynamical system near a bifurcation ensures stochastic switching between different attractors, endowing the simulation with the kind of various firing patterns  seen in the empirical data.

Some other researchers have already  explored deterministic chaos as an alternative to noise-induced multistability to reproduce statistical observables computed from experiment recordings. It is well  known that chaos has a key feature of highly dependent on the initial conditions\cite{strogatz2018nonlinear}. Chaotic itinerancy and  heteroclinic cycling talk about deterministic dynamics\cite{friston2012perception,miller2016itinerancy},endowing  networked chaotic dynamical systems with complex metastable dynamics in the absence of noise.   Previous studies demonstrated that chaos can emerge in medium or large size neural networks\cite{sompolinsky1988chaos,sprott2008chaotic} and even in small circuits or low-dimension  mean-field approximations\cite{ fasoli2016complexity,xu2018synchronization}. Some others studies also  showed  that neuronal activity is intrinsically irregular in the generation of action potentials, their axonal propagation, and the synaptic transmission that follows\cite{faisal2008noise,faisal2005ion}. Therefore, both deterministic and stochastic mechanisms are present in neural dynamics and they should contribute differently to the functional connectivity dynamics. In all, many advances have been made to understand how network topology, connection delays, and noise can contribute to building this dynamic. Little or no attention, however, has been paid to the difference between local chaotic and stochastic influences on the switching between different network states.

More recent has seen a growing body of modelling studies on the comparation and contribution  of  chaos  and noise  to multistable dynamics of neural network. Some earlier studies have explored multistable dynamics of networks composed by chaotic oscillators without noise at all\cite{honey2007network,gollo2014frustrated}, and others have used simple oscillatory or non-oscillatory nodes of their dynamics driven by noise\cite{hansen2015functional,deco2017single}.After that, some works employ  complex dynamic models at the node level but still in presence of neural noise to drive the multistability\cite{pototsky2012emergence,deco2013resting,deco2012ongoing,freyer2011biophysical}. Finally, recent studies have begun to explore the significant difference mechanisms of neural noise and deterministic chaos to drive multistability\cite{orio2018chaos,piccinini2021noise}. Piccinini et al.\cite{piccinini2021noise} showed that chaotic dynamics give rise
to some interesting features in whole-brain activity models, out performing noise-driven equilibrium models in the simultaneous
reproduction of multiple empirical observables.

Our previous study  have shown how channel noise can alter the multistable dynamics that is induced by chaos in a weakly coupled,heterogeneous, network of neural oscillators\cite{orio2018chaos}.Using a conductance-based neural model that can have chaotic dynamics, we found that a network can show multistable dynamics in a certain range of global connectivity strength and under deterministic conditions. We characterized the multistable dynamics when the networks are, in addition to chaotic, subject to ion channel
stochasticity in the form of multiplicative (channel) or additive (current) noise.  More importanly, results of  our previous work shown that moderate noise can enhance the multistable behavior that is evoked by chaos, resulting in more heterogeneous synchronization patterns, while more intense noise abolishes multistability.

At first glance, our results are not as straightforward to interpret as in the previously mentioned works.  With the goal of understanding what role of synaptic noise between excitatory cells might play in influencing collective neural dynamics, we systematically swept the parameters of synaptic noise  and excitatory coupling weights, trying to keep other variables, such as parameters of synaptic dynamics. In this paper, we show that both synaptic  noise intensity and excitatory weights can make a great contribution to the coexistence of firing states  with slight modification parameters. First, we find that neuronal networks can display  time-varying  and parameter-varying multistability.  Second, we  also find that the emergence of  time-varying multistability is shown  as a format  of metstable  state,   while the parameter-varying multistability  is accompanied by coexistence of synchrony state and metastable  (or asynchronous firing state) with slightly varying noise intensity and excitatory weights. Our results offer a series of precise statistical  explanation of the intricate effect of synaptic noise  in neural multistability.

Our findings have some issues that need to be considered when interpreting them.We are studying multistability in a medium-sized network of oscillatory neurons and trying to relate its behavior to phenomena observed at much larger scales. The main difference, which was also explained in previous study\cite{orio2018chaos},  between our model and large scale ones is the nature of synaptic connections between neurons. In our previous studies\cite{xu2013controlling,xu2014simplified,xu2018synchronization,orio2018chaos,xu2021diversity,tian2018chimera,zhou2021synaptic},  we  first have used electrical synapses\cite{xu2018synchronization,orio2018chaos,tian2018chimera,zhou2021synaptic},a type of bidirectional synapse that has a dampening effect with a natural tendency to synchronize a network, while long-range connections in the brain are unidirectional with chemical synaptic  nature. This type of connection generates more complex dynamics, especially when time delays and noise  are considered\cite{deco2009key}. Chemical synaptic coupling is responsible for a variety of network effects, particularly in networks that generate rhythmic activity some of which are well established, such as regulation of phase relationships\cite{xu2013controlling,xu2014simplified}, synchrony\cite{ashhad2020emergent,montbrio2020exact}, and pattern formation\cite{xu2021diversity}, and some are novel, such as  coexistence  of  firing patterns driven by chaos or noise. Recently, many experiments have evidenced that diverse neural circuits use hybrid synapses coexisting in most organisms and brain structures,  the precise mechanisms whereby these hybrid synapses  contribute to  various firing modes of brain activities  are subtle and remain elusive. On the 
other hand,  electrical synapses  are localized primarily in inhibitory neurons widely distributed throughout the CNS, whereas we here presente  a model of  electrical synapses interlinking excitatory neurons\cite{llinas1974electrotonic,hinrichsen1970coupling,nagy2018electrical}.

Brain dynamics is often inherently variable and unstable, consisting of sequences of transient spatiotemporal patterns\cite{sporns2010networks,rabinovich2008transient}. These sequences of transients are a hallmark of metastable dynamics that are neither entirely stable nor completely unstable. The neuronal  noise of brain dynamics, given by synaptic connections  and  channel noise, must play a role in
the observed  multistable dynamics.  Here we present the first attempt to raise the question  and find  two types of coexisting firing patterns.Our results pave a possible way to uncover the underlying mechanisms of  noise-driven coexisting firing patterns, such as  metastable dynamics, that may mediate perception and cognition.

\begin{acknowledgments}
We thank the funding  No.4111190017 (K.Xu) and 4111710001 (M.Zheng) from Jiangsu University(JSU), China. This work was also 
supported by  the National Science Foundation of China under Grant  Nos.12165016 and 12005079.
\end{acknowledgments}

\newpage

\appendix

\begin{widetext}
\section{}

\begin{enumerate}

\item Parameters and functions of Wang-Buzsáki model

\begin{table}[H] 
	{\renewcommand{\arraystretch}{1.4}%
	\begin{tabular}{ p{5cm} p{5cm} p{5cm}  }
	    \hline \\ 
	    $g_{L}=0.1 $ & $g_{Na}=35$ &$g_{K}=9$ ($mS/cm^2$)\\
		$E_{L}=-65$   & $E_{Na}=55$   &$E_{K}=-90$ (mV)\\
		$C_m=1$ ($\mu F)$ &  $\phi = 5$   &  $I_{app}  = 0$ ($\mu A/cm^2$)\\
	\end{tabular}
	
	\begin{tabular}{p{9.5cm} p{5.5cm}}
		$\alpha_{m}(V)=-0.1(V+35)/[exp(-0.1(V+35))+1]$ & $\beta_{m}(V)=4exp[-(V+60)/18]$ \\
		$\alpha_{h}(V)=0.07exp[-(V+58)/20]$ & $\beta_{h}(V)=1/exp[-0.1(V+28)]+1$ \\	
		$\alpha_n(V) = -0.01(V+34)/[exp(-0.1(V+34))-1]$ & $\beta_n(V) = 0.125exp[-(V+44)/80]$ \\
		$ m_{\infty}=\alpha_{m}/(\alpha_{m}+\beta_{m})$ &  $ $  \\
		 &  \\
		\hline
	\end{tabular}}
\caption{ Parameters and functions of Wang-Buzsáki model(1996)\cite{wang1996gamma}}
\label{S1_Table}
\end{table}

\item  The updating rule of chemical synaptic conductances \label{syn:updating}

In simulations and for practical reasons, the updating rule of chemical synaptic conductances is the same as our previous study\cite{xu2021diversity}, following :
 
\begin{align}
\frac{dy}{dt}&={\bar{g}}_{s(E,I,\nu)}f\delta\left(t_{0}+t_{d}-t\right)-\left(\frac{1}{\tau_1}+\frac{1}{\tau_2}\right)y -\frac{g_{syn}}{\tau_1\tau_2} \nonumber \\
\frac{dg_{syn}}{dt}&=y  
\end{align}

\begin{figure}[htp]	
\includegraphics[width=\linewidth]{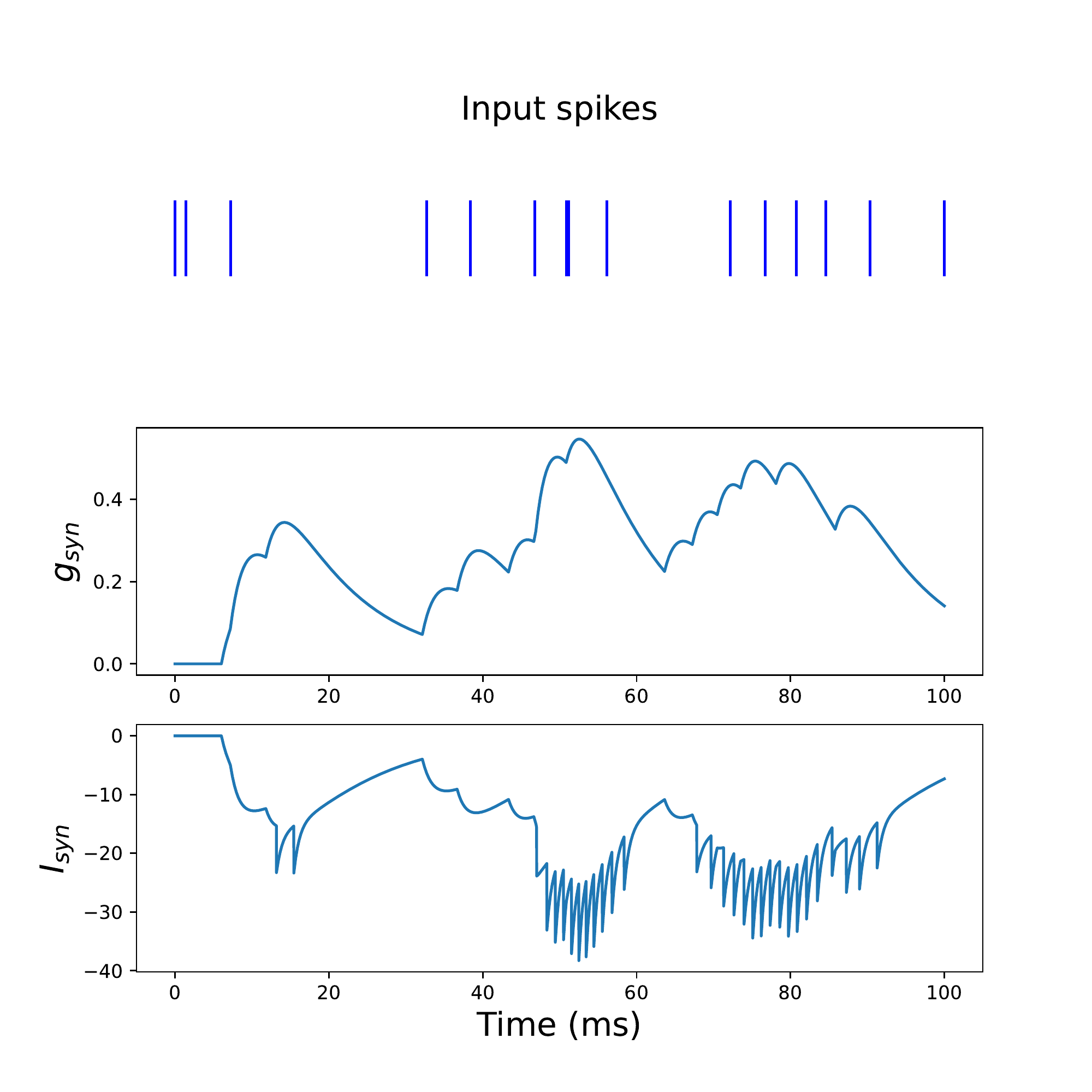}	  
	\caption{Time course of excitatory synaptic input. Top.  presynaptic (input) spikes; Middle.the time course of the synaptic conductance, $g_{syn}$, activated by input spikes; Bottom.excitatory postsynaptic currents (EPSPs) caused  by the simultaneous activation  of synapses.} 
	\label{fig:synapticcurrents}
\end{figure}

The  delta function $\delta\left(t_{0}+t_{d}-t\right)$ represents the spike signal from presynatpic cell, that is $1$ when $t=t_{0}+t_{d}$ and 0 otherwise. The time $t_{0}$ is the time of the presynaptic spike and $t_{d}$ is synaptic delays from the presynatpic to postsynaptic cell.  The conductance peaks occurs at time $t_{peak} = t_{0}+t_{d}+\frac{\tau_{1}\tau_{2}}{\tau_{1}-\tau_{2}}ln(\tau_{1}/\tau_{2})$. The normalization factor $f=1 /(e^{-t_{peak}/\tau_{1}}-e^{-t_{peak}/\tau_{2}})$ and $t_{peak}=\frac{\tau_{1}\tau_{2}}{\tau_{1}-\tau_{2}}ln(\tau_{1}/\tau_{2})$, ensure that the maximum amplitude of the impulse response equals ${\bar{g}}_{s(E,I,\nu)}$. The example  of time course of synaptic conductance and excitatory synaptic currents  activated by input spikes  are shown in Fig.\hyperref[fig:synapticcurrents]{\ref{fig:synapticcurrents}}  In simulations of this work, both peak synaptic conductances ${\bar{g}}_{s(E,I,\nu)}$ and synaptic delays $t_{d}$ are Gaussian distributed random variables with prescribed means $g_{s(E,I)}$ and $\overline{D}$, and standard deviation $\sigma_{g_{s(E,I)}}$ and $\sigma_{\overline{D}}$. The background drive $v_{ext}$ to each neuron in the excitatory population was provided by external excitatory inputs, modeled as independent and identically distributed Poisson processes with rate $\nu$ for different neurons. Peak conductances of the external drive were also heterogeneous and assumed values sampled from a truncated Gaussian distribution with mean $g_{s\nu}$ and standard deviation $\sigma_{g_{s\nu}}$.  These rules are the same as our previous study\cite{xu2021diversity}. Parameters of synaptic model are summarized in  Table  \ref{S2_Table}.

\newpage
\item  List of simulation and default network parameters in the E/I balanced networks 
\begin{table}[H] 
	{\renewcommand{\arraystretch}{1.4}%
	\begin{tabular}{ |p{10cm}| p{2.5cm}| p{2.5cm}|  }
	    \hline
	     \rowcolor{black!80} \multicolumn{3}{l}{\textcolor{white}{A: Global simulation paramters}}  \\ \hline
	    
	    \textbf{Description} & \textbf{Symbol} & \textbf{Value}\\ \hline
	    Simulation duration &  $T_{sim}$ & 2000 ms \\ \hline
	    Start-up transient & $T_{trans}$  & 400 ms \\ \hline
	    Time step  &  dt &    0.02 ms\\ \hline
	    
	    \rowcolor{black!80} \multicolumn{3}{l}{\textcolor{white}{B: Populations and external input}}  \\ \hline
	    \textbf{Description} & \textbf{Symbol} & \textbf{Value}\\ \hline
	    Population size of excitatory neurons& $N_{E}$   & 1000 \\ \hline
	    Population size of inhibitory neurons  & $N_{I}$   & 250 \\ \hline
	    Possion input rate (external excitatory inputs)  &  $\nu$ & 6000 Hz \\ \hline
	    
	    \rowcolor{black!80} \multicolumn{3}{l}{\textcolor{white}{C: Connection parameters}}  \\ \hline
	    \textbf{Description} & \textbf{Symbol} & \textbf{Value}\\ \hline
		Weight of gap junction   & $J_{gap}$   & 0.1 \\ \hline
		Weight of self-excitatory connection   & $J_{EE}$   & 0.05 \\ \hline
		Weight of self-inhibitory connection   & $J_{II}$   & 0.04 \\ \hline
		Weight of inhibitory connections for excitatory population & $J_{EI}$   & 0.03 \\ \hline
        Weight of excitatory connections for inhibitory population  & $J_{IE}$   & 0.01 \\ \hline
        Probability of local inhibitory connections &  $P_{I}$ & 0.2  \\ \hline
        
		\rowcolor{gray!60} \multicolumn{3}{c}{Synaptic dynamics (Difference of Two Exponentials)}  \\ \hline
		Reversal potential for excitatory synapses   & $E_{synE}$   & 0 mV \\ \hline
		Reversal potential for inhibitory synapses   & $E_{synI}$   & -80 mV \\ \hline
		Excitatory synaptic decay time   & $\tau_{1E}$   & 3 ms \\ \hline
		Ihibitory synaptic decay time   & $\tau_{1I}$   & 4 ms \\ \hline
		Excitatory synaptic time & $\tau_{2E}$&  1 ms \\ \hline
		Inhibitory synaptic time & $\tau_{2I}$&  1 ms \\ \hline
		Mean synaptic delay &  $\overline{D}$ &  1.5 ms \\ \hline
		Mean synaptic excitatory conductance &  $g_{sE}$ &  5 nS \\ \hline
		Mean synaptic  inhibitory conductance & $g_{sI}$ & 200 nS \\ \hline
		Mean synaptic  input conductance & $g_{s\nu}$ & 3 nS \\ \hline
		Standard deviation delay & $\sigma_{\overline{D}}$& 0.1 ms \\ \hline
		Standard synaptic excitatory conductance & $\sigma_{g_{sE}}$& 1 nS\\ \hline
        Standard inhibitory conductance & $\sigma_{g_{sI}}$& 10 nS \\ \hline
        Standard input conductance & $\sigma_{g_{s\nu}}$& 1 nS \\ \hline
        		
	    \rowcolor{gray!60} \multicolumn{3}{c}{Small-world networks (SW)}  \\ \hline
	    Probability of replacing new links   &  $P_{sw}$ & 0.01  \\ \hline
		Degree of nearest neighbours connections    & K & 10 \\ 
		\hline
	\end{tabular}}
	\caption{ List of simulation and default network parameters in the E/I balanced networks.}
\label{S2_Table}
\end{table}

\end{enumerate}
\end{widetext}

\nocite{*}

\bibliography{NoiseDriveMulti}

\end{document}